\documentclass[11pt]{article}
\usepackage{amssymb}
\usepackage{amsfonts}
\usepackage{graphicx}
\usepackage{amsmath}

\makeatletter \@addtoreset{equation}{section} \makeatother

\newtheorem{theorem}{Theorem}
\newtheorem{lemma}{Lemma}

\newtheorem{remark}{Remark}
\newtheorem{definition}{Definition}

\def\fracd{\displaystyle\frac}

\begin{document}

\title{Edge  Universality  for Orthogonal Ensembles of Random Matrices}
 \author{ M. Shcherbina\\
 Institute for Low Temperature Physics, Kharkov,
Ukraine. \\E-mail: shcherbi@ilt.kharkov.ua
}
\date{}

\maketitle

\begin{abstract}
We prove edge universality of local eigenvalue statistics  for orthogonal invariant matrix models
with real analytic potentials and  one interval  limiting spectrum.
Our starting point is the result of \cite{S:08} on the representation of the reproducing matrix  kernels
of orthogonal ensembles in terms of scalar reproducing kernel of corresponding unitary ensemble.


\end{abstract}

\section{Introduction and main results}\label{sec:1}

We study ensembles of $n\times n$ real symmetric
(or Hermitian) matrices $M$  with the probability distribution
\begin{equation}
P_{n}(M)dM=Z_{n,\beta}^{-1}\exp \{-\frac{n\beta}{2}\mathrm{Tr}V(M)\}dM,  \label{MMb}
\end{equation}
where $Z_{n,\beta}$ is a normalization  constant, $V:\mathbb{R}\to \mathbb{R%
}_{+}$ is a H\"{o}lder function satisfying the condition
\begin{equation}\label{cond0}
|V(\lambda )|\geq 2(1+\epsilon )\log(1+ |\lambda |).
\end{equation}
A positive parameter $\beta$ here assumes the values $\beta=1$ (in
the case of real symmetric matrices) or $\beta=2$ (in the Hermitian
case), and $dM$ means the  Lebesgue measure on the algebraically
independent entries of $M$.

The
joint eigenvalue distribution corresponding to (\ref{MMb}) has the form (see \cite{Me:91})
\begin{equation}
p_{n,\beta}(\lambda_1,...,\lambda_n)=Q_{n,\beta}^{-1}\prod_{i=1}^n
e^{-n\beta V(\lambda_i)/2}\prod_{1\le j<k\le
n}|\lambda_i-\lambda_j|^\beta, \label{psymb}
\end{equation}
where $Q_{n,\beta}$ is a normalization constant. For both cases ($\beta=1,2$)
 the behavior of  Normalized Counting Measure (NCM) of eigenvalues is now well understood.
 According to
\cite{BPS:95,Jo:98},  NCM converges weakly in probability
 to the non random limiting measure ${\cal N}$ known as
Integrated Density of States (IDS) of the ensemble. The IDS is absolutely continuous,
if $V'$ satisfies the Lipshitz condition  \cite{Sa-To:97}.
The non-negative density $\rho(\lambda) $  is called
 Density of States (DOS) of the ensemble.  IDS can be found as a
unique solution of a certain variational problem (see
\cite{BPS:95,De-Co:98,Sa-To:97}).

To study the local regimes for ensembles (\ref{MMb}) means
to study the behavior of marginal densities
\begin{equation}\label{p_nl}
p^{(n)}_{l,\beta}(\lambda_1,...,\lambda_l)=
\int_{\mathbb{R}^{n-l}} p_{n,\beta}(\lambda_1,...\lambda_l,\lambda_{l+1},...,\lambda_n)
d\lambda_{l+1}...d\lambda_n
\end{equation}
in the scaling limit, when
$\lambda_i=\lambda_0+x_i/n^\kappa$ $(i=1,\dots,l)$,
and $\kappa$ is a constant, depending on the behavior of DOS $\rho(\lambda)$
in a small neighborhood of $\lambda_0$.
If $\rho(\lambda_0)\not=0$, then $\kappa=1$, if $\rho(\lambda_0)=0$
and $\rho(\lambda)\sim |\lambda-\lambda_0|^\alpha$, then $\kappa=1/(1+\alpha)$.
The universality conjecture states that  the scaling limits of all marginal densities
 are universal, i.e. do not depend on $V$ and depend only on $\alpha$ and $\beta$.
One of the most known quantity probing the local regime is the gap probability, i.e.,
the probability that there is no eigenvalues in the interval
$\Delta_n(a,b)=[\lambda_0+a/n^\kappa,\lambda_0+b/n^\kappa]$
\begin{equation}\label{gap}
E_{n,\beta}(\Delta_n(a,b))=\mathbf{E}\bigg\{\prod_{k=1}^k(1-\mathbf{1}_{\Delta_n(a,b)}(\lambda_i))\bigg\}.
\end{equation}
Thus,  results on the universality of local eigenvalue statistics usually include  proofs of
universality of the gap probability.

For unitary ensembles all marginal densities can be represented
 (see \cite{Me:91}))  in terms of so called reproducing kernel
\begin{equation}\label{k_n}
K_{n}(\lambda ,\mu )=\sum_{l=0}^{n-1}\psi _{l}^{(n)}(\lambda )\psi
_{l}^{(n)}(\mu ).
\end{equation}%
where
\begin{equation}\label{psi}
\psi _{l}^{(n)}(\lambda )=\exp \{-n V(\lambda )/2\}p_{l}^{(n)}(\lambda
),\;\,l=0,...,
\end{equation}
and $\{p_{l}^{(n)}\}_{l=0}^n$ are orthogonal polynomials on $\mathbb{R}$
associated with the weight
$ w_{n}(\lambda )=e^{-n V(\lambda )}$, i.e.,
\begin{equation}\label{ortP}
\int p_{l}^{(n)}(\lambda )p_{m}^{(n)}(\lambda )w_{n}(\lambda )d\lambda
=\delta _{l,m}.
\end{equation}%
In particular,
\[E_{n,2}(\Delta_n(a,b))=\det\{1-K_{\Delta_n(a,b)}\},\]
where $\det\{\dots\}$ is  the Fredholm determinant and $K_{\Delta_n(a,b)}$ is the integral operator
with the kernel (\ref{k_n}) in $L^2(\Delta_n(a,b))$.
Hence, the problem to study  marginal distributions is replaced by the problem
to study  the reproducing kernel $K_n(\lambda,\mu)$ in the scaling limit.

The problem was solved in  many cases. For example, in the
 bulk case ($\rho(\lambda_0)\not=0$) it was shown in
\cite{PS:97} (see also \cite{PS:07}) that for a  general class of $V$ (the third
derivative is bounded in the some neighborhood of $\lambda_0$)  the scaled reproducing
kernel converges uniformly to the $\sin$-kernel.
This result for the case of real analytic $V$ was obtained also in
\cite{De-Co:99a}.

  Universality near the edge, i.e., the case when $\lambda_0$ is the edge point of
  the spectrum and $\rho(\lambda)\sim|\lambda-\lambda_0|^{1/2}$, as
$\lambda\sim\lambda_0$, was studied in \cite{De-Co:99a}. It was proved that
\[
\lim_{n\to\infty}\frac{1}{n^{2/3}\gamma}K_{n,2}(\lambda_0+s_1/n^{2/3}\gamma,
\lambda_0+s_2/n^{2/3}\gamma)=Q_{Ai}(s_1,s_2),\]
where
\begin{equation}\label{QA}
Q_{Ai}(s_1,s_2)=\frac{Ai'(s_1)Ai(s_2)-Ai(s_1)Ai'(s_2)}{s_1-s_2}.
\end{equation}
This result for GUE ($V(\lambda)=\lambda^2/2$) was obtained in \cite{Tr-Wi:05}.
There are also results on universality near the extreme
point,  where $\rho(\lambda)\sim (\lambda-\lambda_0)^2$, as
$\lambda\sim\lambda_0$ (see \cite{C-K:06} for  real
analytic $V$ and \cite{S:05} for  general $V$).

 For orthogonal ensembles ($\beta=1$ ) the situation is more complicated. Instead of
(\ref{k_n})
we need to use   the matrix kernel
\begin{equation}\label{K_n1}
\widehat K_{n}(\lambda,\mu)=\left(\begin{array}{cc}S_n(\lambda,\mu)&
S_nd(\lambda,\mu)\\ IS_n(\lambda,\mu)-\epsilon(\lambda-\mu)& S_n(\mu,\lambda)\end{array}\right).
\end{equation}
Here
\begin{equation}\label{S}
S_n(\lambda,\mu)=-\sum_{i,j=0}^{n-1}\psi^{(n)}_i(\lambda)(\mathcal{M}^{(0,n)})^{-1}_{i,j}
(n\epsilon\psi^{(n)}_j)(\mu),
\end{equation}
where $\psi^{(n)}_i$ are defined by (\ref{psi})-(\ref{ortP}) and the matrix $\mathcal{M}^{(0,n)}$
is defined as
\begin{equation}\label{M}
M_{j,l}=n(\psi^{(n)}_j,\epsilon\psi^{(n)}_l);\quad
\mathcal{M}^{(0,\infty)}=\{M_{j,l}\}_{j,l=0}^\infty;\quad
\mathcal{M}^{(0,n)}=\{M_{j,l}\}_{j,l=0}^{n-1},
\end{equation}
where $\epsilon$ is the integral operator with the kernel
\begin{equation}\label{eps}
\epsilon(\lambda)=\frac{1}{2}\hbox{sign}(\lambda);\quad \epsilon
 f(\lambda)=\int\epsilon(\lambda-\mu)f(\mu)d\mu.
\end{equation}
The symbol $d$ in (\ref{K_n1}) denotes the differentiating with respect to $\mu$, and $IS_n(\lambda,\mu)$
means the composition of  operators  $\epsilon$ and $S_n$. Similarly
to the hermitian  case all marginal densities can be
expressed in terms of the kernel $\widehat K_{n}$ (see \cite{Tr-Wi:98}). In particular, the
gap probability has the form
\begin{equation}\label{gap_o}
E_{n,1}(\Delta_n(a,b))=\hbox{det}^{1/2}(1-\widehat K_{n}(\Delta_n(a,b))),
\end{equation}
where $\widehat{K}_{n}(\Delta_n(a,b))$
is an integral operator from
$L^{2}(\Delta_n(a,b))\oplus L^{2}(\Delta_n(a,b))$ to itself defined by the matrix
kernel (\ref{K_n1}) and $\det$ means it Fredholm
determinant.
The matrix kernel  (\ref{K_n1}) was introduced first in
\cite{Dy} for  circular ensemble and then in \cite{Me:91}  for
 orthogonal ensembles. The scalar kernels of (\ref{K_n1}) could
be defined in principle in terms of any family  of polynomials complete in
$L_2(\mathbb{R},w_n)$  (see
\cite{Tr-Wi:98}), but usually the families of skew orthogonal
polynomials were used (see \cite{Me:91} and references therein).
Unfortunately,   using of
skew orthogonal polynomials for general $V$  rises
serious technical difficulties.

The main technical obstacle to study the kernel (\ref{S}) defined
in terms of orthogonal polynomials is that there is no uniform bound
for $||(\mathcal{M}^{(0,n)})^{-1}||$. According to Widom  (see \cite{Wi:99}),
 if the potential $V$ is a rational function, then to control $(\mathcal{M}^{(0,n)})^{-1}$
it is enough to control the inverse of some matrix of fixed size depending of $V$
(e.g. if $V$ is polynomial of degree $2m$, then we should
control some $(2m-1)\times(2m-1)$ matrix).
In the paper \cite{De-G:07}  by constructing of the exact expressions for the entries of the Widom matrix,
it was shown that it is invertible in the  case  $V(\lambda)=\lambda^{2m}$.
 This allowed to prove bulk universality for the case $V(\lambda)=\lambda^{2m}+
 n^{-1/2m} a_{2m-1}\lambda^{2m-1}+\dots$ (in
 our notations). The same approach was used in \cite{De-G:07a} to prove  edge universality and
in \cite{DGKV:07} to prove bulk and edges universality (including the case
of hard edge) for the Laguerre type ensembles with monomial $V$.
 In the papers \cite{St1,St2} universality in the bulk and near the edges   were studied
for  $V$ being an even quatric polynomial.
The most general result for the moment was obtained in \cite{S:08}, where it was  shown that in the one
interval  case  the matrix, which we need to control, is of  rank  one. This allowed to study
any real analytical potential with one interval   support and to simplify considerably the proof.
 In the present paper we will use the result of \cite{S:08} to prove  that up to  some small terms
 (which do not contribute in the limit) the kernel $S$ is the same that for GOE case (see
 Lemma \ref{l:ueo1}). This allows us to use the method of \cite{Tr-Wi:05} to prove universality
 of limiting kernels near the edges. The only difference with \cite{Tr-Wi:05} is that we use asymptotic of
 orthogonal polynomials of \cite{De-Co:99a}, instead of classical asymptotic of Hermite
 polynomials.

Let us state  our main conditions.

\begin{description}
\item[\textbf{ C1.}]\textit{  The support $\sigma $ of
IDS of the ensemble consists of a single interval:}
\[
\sigma =[-2,2].
\]
\item[\textbf{ C2.} ] \textit{$V(z )$ satisfies (\ref{cond0}) and
is an even analytic
function in }
\begin{equation}
\Omega[d_1,d_2]=\{z:-2-d_1\le\Re z\le 2+d_1,\,\, |\Im z|\le d_2\}, \quad
d_1,d_2>0.
\label{Omega}\end{equation}

\item[\textbf{C3.}] \textit{ DOS $\rho(\lambda)$ is
strictly positive in the internal points $\lambda\in (-2,2)$ and $\rho(\lambda)\sim
|\lambda\mp 2|^{1/2}$, as $\lambda\sim\pm2$}.
\item[\textbf{C4.}]\textit{ The function
\begin{equation*}
u(\lambda)=2\int\log |\mu-\lambda|\rho(\mu)d\mu-V(\lambda)
\end{equation*}
 achieves its maximum
if and only if $\lambda\in\sigma$. }
\end{description}

Note (see \cite{APS:01}) that under conditions \textbf{C1-C4} the limiting
density of states (DOS) $\rho $ has the form
\begin{equation}\label{rho}
\rho (\lambda )=\frac{1}{2\pi }P(\lambda
)\sqrt{4-\lambda^2}\,\mathbf{1}_{|\lambda|<2},
\end{equation}
where the function $P$ can be represented in the form
\begin{equation}
P(z)=
\frac{1}{2\pi }\int_{-\pi}^{\pi}{\frac{V^{\prime }(z)-V^{\prime
}(2\cos y )}{z-2\cos y}}dy .  \label{P}
\end{equation}%
If $V$ is a polynomial of $2m$th degree, then it is evident that $P(z)$
is a polynomial of $(2m-2)$th degree, and conditions \textbf{C3} guarantee that
\begin{equation}
  \label{posP}
 |P(z)|\le C,\quad z\in\Omega[d_1/2,d_2/2], \quad P(\lambda)\ge\delta>0,\quad
 \lambda\in[-2,2].
\end{equation}

We will use also that under conditions \textbf{C1-C4} the entries of
  semi infinite Jacoby matrix
$\mathcal{J}^{(n)}$, generated by the recursion relations for
 orthogonal polynomials (\ref{ortP})
\begin{equation}\label{rec}
J_{l+1}^{(n)}\psi _{l+1}^{(n)}(\lambda )+q_{l}^{(n)}\psi _{l}^{(n)}(\lambda )+
J_{l}^{(n)}\psi _{l-1}^{(n)}(\lambda )=\lambda \psi _{l}^{(n)}(\lambda
),\quad J_{0}^{(n)}=0,\quad l=0,....
\end{equation}%
satisfy the relations  (see \cite{APS:97, APS:01}):
  $q^{(n)}_l=0$ and
\begin{equation}\label{J_k}
  \bigg|J^{(n)}_{n+k}-1-\frac{k}{2nP(2)}\bigg|\le C\frac{|k|^2+n^{2/3}}{n^{2}},
  \quad |k|\le 2n^{1/2},
\end{equation}
where $P$ is defined by (\ref{P}).
Here and everywhere below we denote by $C,C_0,C_1,c,...$ positive $n$-independent constants (different
in  different formulas).

The main result of the paper is
\begin{theorem}
\label{t:ueo} Consider an orthogonal ensemble of random matrices
  (\ref{psymb}) with $\beta =1$,  and $V$ satisfying conditions \textbf{C1-C4}.
  Set $\gamma=P^{2/3}(2)$, where $P$ is defined by (\ref{P}). Then for even $n$ we have:

\begin{itemize}
\item[(i)] if $p^{(n)}_{l1}$ is the  $l$th  marginal
of (\ref{psymb}), then $n^{l/3}p^{(n)}_{l1}(2+x_1/\gamma n^{2/3},\dots,2+x_l/\gamma n^{2/3})$ converges uniformly in
$x_{j}\geq s>-\infty ,\;j=1,...,l$ to the limits coinciding with that for GOE
and given in terms of
\begin{equation}
\widehat{Q}_{Ai}(x,y)=\lim_{n\rightarrow \infty }n^{-2/3}\gamma^{-1}\widehat{K}%
_{n}(2+x/\gamma n^{2/3},2+y/\gamma n^{2/3}),  \label{edge}
\end{equation}%
\begin{equation}
\widehat{Q}_{Ai}(x,y)=\left(
\begin{array}{cc}
S_{Ai}(x,y) & D_{Ai}(x,y) \\
I_{Ai}(x,y)-\epsilon (x-y) & S_{Ai}(y,x)%
\end{array}%
\right)  \label{K_A}
\end{equation}%
with
\begin{eqnarray}  \label{S_A}
S_{Ai}(x,y)&=&Q_{Ai}(x,y)+\dfrac{1}{2}Ai(x)\left( 1-\displaystyle%
\int_{y}^{\infty }Ai(z)dz\right) , \\
D_{Ai}(x,y)&=&-\partial _{y}Q_{Ai}(x,y)-\dfrac{1}{2}Ai(x)Ai(y),  \notag \\
I_{Ai}(x,y)&=&-\displaystyle\int_{x}^{\infty }Q_{Ai}(z,y)dy  \notag \\
\hspace{0.5cm}&&+\frac{1}{2}\left( \int_{x}^{y}Ai(z)dz+\int_{x}^{\infty
}Ai(z)dz\int_{y}^{\infty }Ai(z)dz\right) ,  \notag
\end{eqnarray}%
and $Q_{Ai}(x,y)$ of (\ref{QA});

\item[(ii)] \ if $E_{n,1}$ is the gap probability (\ref{gap_o}) of  (\ref{psymb}),
corresponding to the semi-infinite interval $\left( 2+s/\gamma n^{2/3},\infty
\right) $, then
\begin{equation}
\lim_{n\rightarrow \infty }E_{n,1}\left( \left( 2+s/\gamma n^{2/3},\infty
\right) \right): =E_{1}^{(edge)}(s)=\emph{det}^{1/2}(I-\widehat{Q}_{Ai}(s)),  \label{E1soft}
\end{equation}%
where $\widehat{Q}_{Ai}(s)$ is the integral operator, defined in $%
L^{2}(s,\infty ;w)\oplus L^{2}(s,\infty ;w^{-1})$ by the $2\times 2$ matrix
kernel (\ref{K_A}), (\ref{S_A}) with $w(x)=x^{2}+1$.
\end{itemize}
\end{theorem}
\begin{remark}\label{r:1}
According to the
results of  \cite{APS:01} and \cite{PS:03}, if we restrict the
integration in (\ref{psymb}) by $|\lambda_i|\le L=2+d_1/2$, consider
the polynomials $\{p^{(n,L)}_k\}_{k=0}^\infty$ orthogonal on the interval
$[-L,L]$ with the weight
$e^{-nV}$ and set $\psi^{(n,L)}_k=e^{-nV/2}p^{(n,L)}_k$,
then for $k\le n(1+\varepsilon)$ with some $\varepsilon>0$
\begin{equation}\label{apr_pol}\begin{array}{l}
\sup_{|\lambda|\le L}|\psi^{(n,L)}_k(\lambda)-
\psi^{(n)}_k(\lambda)|\le e^{-nC},\quad
|\psi^{(n)}_k(\pm L)|\le e^{-nC}
\end{array}\end{equation}
 with some
absolute $C$. Therefore  from the very beginning we can take all
integrals in (\ref{psymb}), (\ref{ortP}), (\ref{eps}) and (\ref{M})
over the interval $[-L,L]$. Note also that
since $V$ is an analytic function in $\Omega[d_1,d_2]$ (see
(\ref{Omega})), for any  $m\in\mathbb{N}$ there exists a polynomial
$V_m$ of the $(2m)$th degree such that
\begin{equation}\label{aprV}
|V_m(z)|\le C_0,\quad |V(z)-V_m(z)|\le e^{-Cm},\quad z\in\Omega[d_1/2,d_2/2].
\end{equation}

Take
\begin{equation}\label{m=}
m=[\log^2 n]
\end{equation}
and consider the system of polynomials
$\{p^{(n,L,m)}_k\}_{k=0}^\infty$  orthogonal in the interval $[-L,L]$
with respect to the weight $e^{-nV_m(\lambda)}$. Set
$\psi^{(n,L,m)}_k=p^{(n,L,m)}_ke^{-nV_m/2}$ and construct
$\mathcal{M}^{(0,n)}_m$ by (\ref{M}) with $\psi^{(n,L,m)}_k$.
Then for any $k\le n+2n^{1/2}$ and uniformly in $\lambda\in[-L,L]$
\begin{equation}\label{dif_M}\begin{array}{l}
|\psi^{(n,L)}_k(\lambda)-\psi^{(n,L,m)}_k(\lambda)|
\le e^{-C\log^2n},\quad |\varepsilon\psi^{(n,L)}_k(\lambda)-\varepsilon\psi^{(n,L,m)}_k(\lambda)|
\le e^{-C\log^2n}\\
 ||\mathcal{M}^{(0,n)}_m-\mathcal{M}^{(0,n)}||\le e^{-C\log^2n},\quad
 ||(\mathcal{M}^{(0,n)}_m)^{-1}-(\mathcal{M}^{(0,n)})^{-1}||\le e^{-C\log^2n}.
\end{array}\end{equation}
The proof of the first bound here is identical to the proof of (\ref{apr_pol})
(see \cite{PS:03}). The second bound follows from the first one because the operator
$\varepsilon:L_2[-L,L]\to C[-L,L]$ is bounded by $L$. The third bound in
(\ref{dif_M}) follows from the first, and the last bound follows from the third one and from
the fact that $||(\mathcal{M}^{(0,n)}_m)^{-1}||$ is uniformly bounded (see \cite{S:08}). Hence
\begin{equation}\label{aprM}\begin{array}{l}
|S_{n,m}(\lambda,\mu)-S_{n}(\lambda,\mu)|\le Cn^4 e^{-C\log^2n}\le e^{-C'\log^2n},
\end{array}\end{equation}
and below we will study  $S_{n,m}(\lambda,\mu)$ instead of
 $S_{n}(\lambda,\mu)$. To simplify notations we omit the indexes
$m,L$, but keep the dependence on $m$ in the estimates. For more detail of the replacement see \cite{S:08}.
\end{remark}
\smallskip

Our starting point is the representation  of the matrix $(\mathcal{M}^{(0,n)})^{-1}$
valid under conditions \textbf{C1-C4} (see Corollary 1 in \cite{S:08}).
 To formulate this result we  introduce a few Toeplitz matrices.
Consider the infinite matrix  $\mathcal{P}=\{P_{j,k}\}_{j,k=-\infty}^\infty$ in $l_2[-\infty,\infty]$
with entries
\begin{equation}\label{calP}
 P_{j,k}=\frac{1}{2\pi}\int_{-\pi}^\pi P(2\cos y)e^{i(j-k)y}dy,
\end{equation}
and  $\mathcal{R}=\mathcal{P}^{-1}$,
\begin{equation}\label{R}
\mathcal{R}^{(0,n)}=\{R_{j,k}\}_{j,k=0}^{n-1}\quad
 R_{j,k}=R_{j-k}=\frac{1}{2\pi }\int_{-\pi}^{\pi}\frac{e^{i(j-k)x}dx}{P(2\cos x)}.
\end{equation}
It is important  that
(see \cite{S:08}), Proposition 1)
\begin{equation}\label{exp_dec}
|R_{j,k}|\le e^{-c|j-k|},\quad|((\mathcal{R}^{(0,n)}))^{-1}_{j,k}|\le e^{-c|j-k|},
\end{equation}
where $c>0$ is some $n$-independent constant.
Remark also that if we denote by $\mathcal{J}^*$ an infinite Jacobi matrix with  constant coefficients
\begin{equation}\label{J^*}
 \mathcal{J}^{*}=\{J^*_{j,k}\}_{j,k=-\infty}^{\infty},\quad J^*_{j,k}=\delta_{j+1,k}
 +\delta_{j-1,k},
\end{equation}
then the spectral theorem yields that $\mathcal{P}=P(\mathcal{J}^*)$,
$\mathcal{R}=P^{-1}(\mathcal{J}^*)$.

Two more  matrices which we use below have the form
\begin{equation}\label{d^*}
\mathcal{D}^{(0,n)}=\{D_{j,k}\}_{j,k=0}^{n-1},\quad
 D_{j,k}=\delta_{j+1,k}-\delta_{j-1,k}.
\end{equation}
and $\mathcal{V}^{(0,\infty)}=\{\mathcal{V}_{j,l}\}_{j,l=0}^\infty,$
where
\begin{equation}\label{calV}\mathcal{V}_{j,l}=\hbox{ sign}(l-j)(\psi^{(n)}_j,V'\psi^{(n)}_l)_2=
\frac{2}{n}\left\{\begin{array}{ll}
(\psi^{(n)}_j,
(\psi^{(n)}_l)')_2,&j>l,\\
(\psi^{(n)}_j,
(\psi^{(n)}_l)')_2+O(e^{-C\log^2n}),&j\le l.
\end{array}\right.\end{equation}
Here $O(e^{-C\log^2n})$ appears because of the integration by parts and bounds (\ref{apr_pol}),
(\ref{dif_M}).

According to Corollary 1 from \cite{S:08}, under conditions \textbf{C1-C4}
\begin{equation}\label{M^-1}
  (\mathcal{M}^{(0,n)})^{-1}_{j,k}=
\mathcal{Q}^{(0,n)}_{j,k}
  +\frac{1}{2}a_jb_k+O(n^{-1/2}\log^6n),\end{equation}
where
\begin{equation}\label{Q}
\mathcal{Q}^{(0,n)}_{j,k}=\frac{1}{2}\left\{\begin{array}{lll}
\mathcal{V}^{(0,\infty)}_{j,k},&\,\hbox{for}\,\,\,\,& 0\le j\le n-2m,\,0\le k<n,
\\ (\mathcal{R}^{(0,n)})^{-1}\mathcal{D}^{(0,n)})_{j,k},& \, \hbox{for}\,\,\,\,
& n-2m< j< n,\,0\le k<n.
\end{array}\right.
\end{equation}
 and
\begin{equation}\label{a,b}
a_j=((\mathcal{R}^{(0,n)})^{-1}e_{n-1})_{j},\quad
 b_k=((\mathcal{R}^{(0,n)})^{-1}r^*)_{k},\quad
r^*_{n-i}=R_i
\end{equation}
 with $R_i$ defined by (\ref{R}).

Note that since $(\mathcal{R})^{-1}_{j,k}=\mathcal{P}_{j,k}=0$ for $|j-k|>2m-2$,
the standard linear algebra  yields that $(\mathcal{R}^{(0,n)})^{-1}$ possesses the same
property, i.e.,
\begin{equation}\label{pr_R}
(\mathcal{R}^{(0,n)})^{-1}_{j,k}=0,\, \hbox{for}\, |j-k|>2m-2\Rightarrow
\mathcal{Q}^{(0,n)}_{j,k}=0,\, \hbox{for}\, |j-k|>2m-2.
\end{equation}

\section{Proof of Theorem \ref{t:ueo}.}

\begin{remark}
\label{r:t_leo1} \textit{(1.)} It is easy to see that the integral operator
with the kernel $\widehat{K}_{n}(2+x/\gamma n^{2/3},\lambda _{0}+y/\gamma n^{2/3})$
defined in (\ref{K_n1})  is not a trace class operator in $%
L^{2}(s,\infty )\oplus L^{2}(s,\infty )$  (recall
that  $A:\mathcal{H}%
_{1}\rightarrow \mathcal{H}_{2}$ is a trace class operator if $||A||_{1}:=\hbox{Tr}\,(A^{\ast
}A)^{1/2}<\infty$).
To take care of this problem we
follow the approach of \cite{Tr-Wi:05} and use weighted $L^{2}$ spaces. If
we take any $w$ such that $w^{-1}\in L_{1}$ and grows at infinity not
faster than exponentially, then $\widehat{K}_{n}$ is a Hilbert-Schmidt on $%
L^{2}(s,\infty ;w)\oplus L^{2}(s,\infty ;w^{-1})$. The diagonal entries of $%
\widehat{K}_{n}$ are finite rank, hence trace class. Now the definition of
determinant extends to Hilbert-Schmidt operator matrices $\widehat{T}$ with
trace class diagonal entries by setting
\begin{equation*}
\det (I-\widehat{T})=\det\nolimits_{2}(I-\widehat{T})e^{-\hbox{Tr\,}\widehat{%
T}},
\end{equation*}%
where $\hbox{Tr\,}$ denotes the sum of the traces of the diagonal entries of
$\widehat{T}$ and the $\det\nolimits_{2}$ is the regularized 2-determinant,
defined for the Hilbert-Schmidt operator $T$ with eigenvalues $t _{k}$ as
\[
\det\nolimits_{2}(I-\widehat{T})=\prod (1-t _{k})e^{t _{k}}\]
(see \cite{Go-Kr:69}, Section IV.2). It follows from the identity for 2-determinant
\begin{equation*}
\det\nolimits_{2}(I-\widehat{T}_{1})(I-\widehat{T}_{2})e^{\hbox{Tr\,}%
\widehat{T}_{1}\widehat{T}_{2}}=\det\nolimits_{2}(I-\widehat{T}%
_{1})\det\nolimits_{2}(I-\widehat{T}_{2})
\end{equation*}%
that for this extended definition we still have the relation
\begin{equation*}
\det (I-\widehat{T}_{1})(I-\widehat{T}_{2})=\det (I-\widehat{T}_{1})\det (I-%
\widehat{T}_{2})
\end{equation*}%

\textit{(2.)} Consider a rank one kernel $u(x)v(y)$, where $u\in
L^{2}(s,\infty ;w_2)$ and $v\in L^{2}(s,\infty ;w_1^{-1})$ and
  $u\otimes v:L^{2}(s,\infty ;w_{1})\rightarrow
L^{2}(s,\infty ;w_{2})$
\begin{equation}\label{intr1}
(u\otimes v\,h)(x)=u(x)\int_{s}^{\infty }h(y)v(y)dy.
\end{equation}%
Then we have
\begin{equation}
||u\otimes v||_{1}\leq ||u||_{L^{2}(w_{2})}||v||_{L^{2}(w_{1}^{-1})}
\label{||uv||}
\end{equation}%

\end{remark}

Introduce the scaled kernels:
\begin{equation}
\begin{array}{rcl}
\mathcal{S}_{n}(x,y) & = & (n^{2/3}\gamma)^{-1}S_{n}(2+x/\gamma n^{2/3},2+y/\gamma n^{2/3}), \\
\mathcal{D}_{n}(x,y) & = & (n^{2/3}\gamma)^{-2}D_{n}(2+x/\gamma n^{2/3},2+y/\gamma n^{2/3}), \\
\mathcal{I}_{n}(x,y) & = & I_{n}(2+x/\gamma n^{2/3},2+y/\gamma n^{2/3}), \\
\mathcal{K}_{n}(x,y) & = & (n^{2/3}\gamma)^{-1}K_{n}(2+x/\gamma n^{2/3},2+y/\gamma n^{2/3}),%
\end{array}
\label{cal_S}
\end{equation}%
where $S_{n}$ is defined by (\ref{S}) and $K_{n}$ is defined
in (\ref{k_n}).

We prove first assertion (ii). Observe that the determinant in (\ref{gap_o})
is the same if we replace the interval $(2+s/n^{2/3},2+\varepsilon )$ by $(s,\varepsilon n^{2/3})$
 and the kernel $\widehat{K}_{n}$ by
\begin{equation}
\widehat{\mathcal{K}}_{n}(x ,y )=\left(
\begin{array}{cc}
\mathcal{S}_{n}(x,y) & \mathcal{D}_{n}(x,y) \\
\mathcal{I}_{n}(x,y)-\epsilon (x-y) & \mathcal{S}_{n}(y,x)%
\end{array}%
\right).  \label{cal_K_1}
\end{equation}%

\begin{lemma}\label{l:ueo1}
If we denote
\begin{equation}\label{lueo.1}
\varphi _{n}(x)=n^{-1/6}\gamma^{-1/2}\psi _{n}^{(n)}(2+x/\gamma n^{2/3}),\quad \psi
_{n}(x)=n^{-1/6}\gamma^{-1/2}\psi _{n-1}^{(n)}(2+x/\gamma n^{2/3}),
\end{equation}%
then
\begin{eqnarray}
\mathcal{S}_{n}(x,y)&=&\mathcal{K}_{n}(x,y)+\frac{1}{2}\psi _{n}(x)\epsilon \varphi
_{n}(y)+r_n(x,y)\notag\\
\mathcal{D}_{n}(x,y)&=&-\frac{\partial}{\partial y}\mathcal{K}_{n}(x,y)-\frac{1}{2}\psi _{n}(x) \varphi
_{n}(y)-\frac{\partial}{\partial y}r_n(x,y)\label{rep_S}\\
\mathcal{I}_{n}(x,y)&=&\mathcal{I}\mathcal{K}_{n}(x,y)+\frac{1}{2}\epsilon \psi _{n}(x)\epsilon \varphi
_{n}(y)+(\epsilon r_n)(x,y)\notag
\end{eqnarray}
where
\begin{equation}\label{lueo.2}
||r_n(x,y)||_1,\,||\frac{\partial}{\partial y}r_n(x,y)||_1,\,
||(\epsilon r_n)(x,y)||_1\le Cn^{-1/3}\log^6 n.
\end{equation}
\end{lemma}
\textbf{Proof}.
Since  (\ref{M^-1}), (\ref{Q}) and (\ref{calV}) imply  for $j\le n-2m$
\[n\sum_{k=0}^{n-1}(\mathcal{M}^{(0,n)})^{-1}_{j,k}\epsilon\psi^{(n)}_k(\mu)=-\psi^{(n)}_j(\mu)
+O(e^{-c\log^2n}),\]
we have
\begin{equation}\label{lueo.3}
-n\sum_{j=0}^{n-2m}\sum_{k=0}^{n-1}\psi^{(n)}_j(\lambda)(\mathcal{M}^{(0,n)})^{-1}_{j,k}\epsilon\psi^{(n)}_k(\mu)
=\sum_{j=0}^{n-2m}\psi^{(n)}_j(\lambda)\psi^{(n)}_j(\mu)+O(e^{-c\log^2n}).
\end{equation}
For $n-1\le j> n-2m$ we need to use the result of \cite{S:08} (see Eq. (68)), according to which for any $|p-n|\le 4\log^2
n$ we have
\begin{multline}\label{d_eps}
-\frac{1}{2}\left(\epsilon\psi^{(n)}_{p+1}(\mu)-\epsilon\psi^{(n)}_{p-1}(\mu)\right)=n^{-1}\sum_{l=0}^{\infty}
R_{p,l}\psi^{(n)}_l(\mu)+n^{-1}e_p(\mu) \\=
n^{-1}\sum_{l=0}^{n-1}
R_{p,l}^{(0,n)}\psi^{(n)}_l(\mu)+n^{-1}\sum_{l=n}^{\infty}
R_{p,l}\psi^{(n)}_l(\mu)+n^{-1}e_p(\mu),
\end{multline}
where the remainder terms $e_p(\mu)$ satisfy the bounds
\[||e_p||_{L^2[-L,L]}\le Cn^{-1/2}\log^4 n.\]
Therefore the scaled functions $\widetilde e_p(x)=n^{-1/3}e_p(2+x/\gamma n^{2/3})$ admit the bounds
\[||\widetilde e_p||_{L^2(w^{-1})}\le Cn^{-1/2}\log^4 n.\]
Hence, using  the definition of $\mathcal{D}^{(0,n)}$ (\ref{d^*}) and (\ref{d_eps}), we obtain
\begin{multline}\label{lueo.4}
-n\sum_{j=n-2m+1}^{n-1}\sum_{k=0}^{n-1}\psi^{(n)}_j(\lambda)
\left((\mathcal{R}^{(0,n)})^{-1}\mathcal{D}^{(0,n)}\right)_{j,k}\epsilon\psi^{(n)}_k(\mu)
=\sum_{j=n-2m+1}^{n-1}\psi^{(n)}_j(\lambda)\psi^{(n)}_j(\mu)\\
+\frac{n}{2}\epsilon\psi^{(n)}_n(\mu)\sum_{j=n-2m+1}^{n-1}\psi^{(n)}_j(\lambda)(\mathcal{R}^{(0,n)})^{-1}_{j,n-1}+
r_n^{(1)}(\lambda,\mu)+r_n^{(2)}(\lambda,\mu),
\end{multline}
where $r_n^{(1)}(\lambda,\mu)$ collects the terms, which appear because of the second sum in the r.h.s. of
(\ref{d_eps}), and $r_n^{(2)}(\lambda,\mu)$ collects the remainder terms $e_p$ of (\ref{d_eps}):
\begin{eqnarray*}
r_n^{(1)}(\lambda,\mu)&:=&\sum_{j=n-2m+1}^{n-1}\sum_{p=0}^{n-1}\sum_{l=n}^{\infty}
(\mathcal{R}^{(0,n)})^{-1}_{j,p}R_{p,l}
\psi^{(n)}_j(\lambda)\psi^{(n)}_l(\mu),\\
r_n^{(2)}(\lambda,\mu)&:=&\sum_{j=n-2m+1}^{n-1}\sum_{p=0}^{n-1}(\mathcal{R}^{(0,n)})^{-1}_{j,p}
\psi^{(n)}_j(\lambda)e_p(\mu).
\end{eqnarray*}
\begin{definition}
We will say that some remainder kernel $r^{(\alpha)}_n(\lambda,\mu)$
($\alpha=1,2,\dots$)  satisfies the bound \textbf{B} with exponents $\kappa_1$, $\kappa_2$ and $\kappa_3$, if
\[
   || n^{-2/3}r_n^{(\alpha)}(2+x/\gamma n^{2/3},2+y/\gamma n^{2/3})||_1
   \le Cn^{-\kappa_1}m^{\kappa_2}\log^{\kappa_3} n. \eqno(\emph{\textbf{B}})
\]
\end{definition}
According to the results of \cite{De-Co:99a}, we have
\begin{eqnarray}\label{as_De}
n^{-1/6}\gamma^{-1/2}\psi^{(n)}_j(2+x/\gamma n^{2/3})&=&Ai(x+(n-j)/c_* n^{1/3})(1+O(n^{-1/3})),\\
|n^{-1/6}\gamma^{-1/2}\psi^{(n)}_j(2+x/\gamma n^{2/3})|&\le& Ce^{-x},\notag
\end{eqnarray}
where $c_*$ is some constant not important for us. These asymptotic implies, in particular,
that
\begin{eqnarray}\label{as_De.1}
&||n^{-1/6}\gamma^{-1/2}\psi^{(n)}_j(2+x/\gamma n^{2/3})||_{L^2(w)}\le C,\notag\\
&||n^{-1/6}\gamma^{-1/2}\psi^{(n)}_j(2+x/\gamma n^{2/3})||_{L^2(w^{-1})}\le C,
\end{eqnarray}
Using the asymptotic, (\ref{exp_dec}) and (\ref{||uv||}), we obtain that $r_n^{(1)}(\lambda,\mu)$ satisfies (\textbf{B})
with $\kappa_1=1/3$, $\kappa_2=1$, $\kappa_3=0$.

Similarly, using  (\ref{d_eps}), (\ref{as_De}),  (\ref{exp_dec}) and (\ref{||uv||}), we get
that $r_n^{(2)}(\lambda,\mu)$ satisfies (\textbf{B})
with $\kappa_1=2/3$, $\kappa_2=1$, $\kappa_3=4$.

Moreover,  if we denote
\begin{equation}\label{r3}
r_n^{(3)}(\lambda,\mu)=n\epsilon\psi^{(n)}_n(\mu)
\sum_{j=n-2m+1}^{n-1}(\psi^{(n)}_j(\lambda)-\psi^{(n)}_{n-1}(\lambda))(\mathcal{R}^{(0,n)})^{-1}_{j,n-1}
\end{equation}
then (\ref{lueo.3}) and ({\ref{lueo.4}) give us
\begin{multline}\label{lueo.5}
-n\sum_{j=0}^{n-1}\sum_{k=0}^{n-1}\psi^{(n)}_j(\lambda)(\mathcal{M}^{(0,n)})^{-1}_{j,k}\epsilon\psi^{(n)}_k(\mu)
=K_n(\lambda,\mu)\\
+\frac{n}{2}\epsilon\psi^{(n)}_n(\mu)\psi^{(n)}_{n-1}(\lambda)((\mathcal{R}^{(0,n)})^{-1}e_{n-1},u)
+r_n^{(1)}(\lambda,\mu)+r_n^{(2)}(\lambda,\mu)+r_n^{(3)}(\lambda,\mu),
\end{multline}
where $u$ is a vector, whose components are given by
\begin{equation}\label{u}
 u_i=1,\quad  i\in[n-2m,n-1],\quad u_i=0 ,\quad  i\not\in[n-2m,n-1],
 \end{equation}
and  the remainder term $r_n^{(3)}$ in view of (\ref{as_De}),  (\ref{exp_dec}) and (\ref{||uv||})
satisfies the bound (\textbf{B}) with $\kappa_1=1/3$, $\kappa_2=1$, $\kappa_3=0$.

Now we consider the term (see (\ref{M^-1}))
\[ A(\lambda,\mu)=\frac{n}{2}\sum_{k=n-2m}^{n-1}
a_k\psi^{(n)}_{k}(\lambda)\sum_{j=n-2m}^{n-1}b_j\epsilon\psi^{(n)}_j(\mu).\]
Using (\ref{d_eps}) and (\ref{as_De}), similarly to the above it is easy to obtain
that
\begin{equation}\label{A}
 A(\lambda,\mu)=\frac{n}{2}\,(a,u)(b,u)\,\psi^{(n)}_{n-1}(\lambda)\,\epsilon\psi^{(n)}_n(\mu)
 +r_n^{(4)}(\lambda,\mu),\end{equation}
where $u$ is defined in (\ref{u}),
and the remainder $r_n^{(4)}$ satisfies the bound (\textbf{B})
with $\kappa_1=1/3$, $\kappa_2=0$, $\kappa_3=0$.

Let us find $(a,u)(b,u)$. Making transposition  in (\ref{M^-1}) and taking into account
that $(\mathcal{M}^{(0,n)})^{-1}$ and $\mathcal{D}^{(0,n)}$ are skew symmetric matrices, we get
\[-(\mathcal{M}^{(0,n)})^{-1}_{j,k}=-\frac{1}{2}(\mathcal{D}^{(0,n)}(\mathcal{R}^{(0,n)})^{-1})_{j,k}+
\frac{1}{2}a_kb_j+O(n^{-1/2}\log n).\]
Taking the sum of the equation with (\ref{M^-1}) and applying the result to $u$ we get
\[(a,u)(b,u)=\frac{1}{2}([\mathcal{D}^{(0,n)},(\mathcal{R}^{(0,n)})^{-1}]u,u)=
-((\mathcal{R}^{(0,n)})^{-1}u,\mathcal{D}^{(0,n)}u)+O(n^{-1/2}m^2\log n).\]
But it is easy to see that
\[\mathcal{D}^{(0,n)}u=-e_{n-1}+e_{n-2m}+e_{n-2m-1}\]
Hence,
\[
(a,u)(b,u)=((\mathcal{R}^{(0,n)})^{-1}e_{n-1},u)-((\mathcal{R}^{(0,n)})^{-1}(e_{n-2m}+e_{n-2m-1}),u)
+O(n^{-1/2}m^2\log n).
\]
Moreover, since $\mathcal{P}=\mathcal{R}^{-1}$ has only $2m-2$ nonzero diagonals,
the standard linear algebra argument yields that for $j\le n-2m$
$(\mathcal{R}^{(0,n)})^{-1}e_j=\mathcal{P}e_j$. Then, using (\ref{calP}) and (\ref{u}),  we obtain
\[(\mathcal{P}(e_{n-2m}+e_{n-2m-1}),u)=P(2)\]
Finally
\begin{equation}\label{(a,u)(b,u)}
(a,u)(b,u)=((\mathcal{R}^{(0,n)})^{-1}e_{n-1},u)-P(2)+O(n^{-1/2}m^2\log n).
\end{equation}
Then, combining (\ref{lueo.5}) with (\ref{(a,u)(b,u)}) and bounds (\textbf{B})
for $r_n^{(\alpha)}(\lambda,\mu)$ with $\alpha=1,2,3,4$, we get the first line of (\ref{rep_S}). The
second
line of (\ref{rep_S}) can be proved similarly, if we use that (\ref{as_De}) can be differentiated.
To prove the last line of (\ref{rep_S}) we used that (\ref{as_De}) implies that
for $|k-n|=o(n)$
\begin{equation}\label{as_De.2}
    |\epsilon\psi_k(\mu)|\le Cn^{-1/2}.
\end{equation}
Hence
\begin{equation}\label{as_De.3}
    ||\epsilon\psi_k(2+x/\gamma n^{2/3})||_{L^2(w^{-1})}\le Cn^{-1/2}.
\end{equation}
Using   these bounds we get the estimates for
$\epsilon r^{(1)}(\lambda,\mu)$ and $\epsilon r^{(2)}(\lambda,\mu)$. The bound for $\epsilon r^{(3)}(\lambda,\mu)$
and $\epsilon r^{(4)}(\lambda,\mu)$
follow from (\ref{d_eps}) (\ref{||uv||}), (\ref{as_De.1}) and (\ref{exp_dec}).

$\square$

Let us transform the kernel $K_n$. We  use the representation
\begin{equation}\label{repK}
  K_n(\lambda,\mu)=\frac{n}{2}\sum_{k=0}^{n-1}\sum_{j=n}^{\infty}V'(\mathcal{J}^{(n)})_{j,k}\int_0^\infty
  d\nu\left(
  \psi^{(n)}_{k}(\lambda+\nu)\psi^{(n)}_{j}(\mu+\nu)
  +\psi^{(n)}_{k}(\mu+\nu)\psi^{(n)}_{j}(\lambda+\nu)\right).
\end{equation}
The representation can be obtained by taking $\fracd{\partial}{\partial \lambda}+\fracd{\partial}{\partial \mu}$
from both sides of (\ref{repK}) and using of (\ref{calV}) to expand $\fracd{\partial}{\partial
\lambda}\psi^{(n)}_k(\lambda)$ with respect to the basis $\{\psi^{(n)}_j\}_{j=1}^{\infty}$.

Using  the same trick as above, on the basis of (\ref{as_De}) and (\ref{||uv||}) it is easy to show that
\begin{multline}\label{repK1}
  K_n(\lambda,\mu)=\frac{n}{2}\left(\sum_{k=0}^{n-1}\sum_{j=n}^{\infty}V'(\mathcal{J}^{(n)})_{j,k}\right)
  \int_0^\infty
  d\nu\left(
  \psi^{(n)}_{n}(\lambda+\nu)\psi^{(n)}_{n-1}(\mu+\nu)\right.\\ \left.
  +\psi^{(n)}_{n}(\mu+\nu)\psi^{(n)}_{n-1}(\lambda+\nu)\right)+ r_n^{(5)}(\lambda,\mu).
\end{multline}
where $r_n^{(5)}(\lambda,\mu)$ satisfies (\textbf{B}) with $\kappa_1=1/3$, $\kappa_2=\kappa_3=0$.

Moreover, using (\ref{J_k}), we have
\[V^s_n:=\sum_{k=0}^{n-1}\sum_{j=n}^{\infty}V'(\mathcal{J}^{(n)})_{j,k}=
\sum_{k=0}^{n-1}\sum_{j=n}^{\infty}V'(\mathcal{J}^{*})_{j,k}+O(n^{-1})=
\sum_{k=1}^\infty kV'_k+O(n^{-1}),
\]
where
\[V'_k=\frac{1}{2\pi}\int_{-\pi}^\pi V'(2\cos x)e^{ikx}dx.\]
On the other hand, it is evident that if we consider
\[\mathcal{V}(x)=\sum_{k=0}^{n-1}V'_k\sin kx,\]
then
\[V^s_n=\frac{d}{dx}\mathcal{V}(x)\bigg|_{x=0}+O(n^{-1}).\]
But it was proved in \cite{S:08} (see Lemma 1) that $\mathcal{V}(x)=\sin xP(2\cos x)$. Thus we get
\[V^s_n=P(2)+O(n^{-1}),\]
and  we obtain from (\ref{repK1}) that the kernel $\mathcal{K}_n$ from (\ref{cal_S})
can be represented in the form
\begin{equation}\label{repK2}
\mathcal{K}_n(x,y)=\frac{1}{2}  \int_0^\infty
  d z\left(
  \psi_{n}(x+z)\phi_{n}(y+z)+\psi_{n}(y+z)\phi_{n}(x+z)\right)+  r_n^{(6)}(x,y),
\end{equation}
where $||r_n^{(6)}||_1\le Cn^{-1/3}$. Hence, the kernel $\mathcal{S}_n$ is represented
in the form
\begin{multline}\label{repS}
    \mathcal{S}_n(x,y)=\frac{1}{2}  \int_0^\infty
  d z\left(
  \psi_{n}(x+z)\phi_{n}(y+z)+\psi_{n}(y+z)\phi_{n}(x+z)\right)\\
  +\frac{n}{2}\psi_n(x)\epsilon\varphi_n(y)+r_n(x,y),\quad ||r_n||_1\le Cmn^{-1/3}.
\end{multline}
Thus we can prove that $\mathcal{S}_n$ converges in the trace norm to $S_{Ai}$, repeating
almost literally argument of \cite{Tr-Wi:05}, but using (\ref{as_De}) instead of classical asymptotic
for Hermite polynomials. Indeed,
relations (\ref{as_De}) yield
\begin{equation}
\lim_{n\rightarrow \infty }||\varphi _{n}(\cdot +z)-Ai(\cdot
+z)||_{L^{2}(w)}=\lim_{n\rightarrow \infty }||\psi _{n}(\cdot +z)-Ai(\cdot
+z)||_{L^{2}(w)}=0.  \label{TW34a}
\end{equation}%
Let us prove that $\mathcal{K}_{n}:L^{2}(w)\rightarrow L^{2}(w)$ of (\ref%
{cal_S}) converges in the $||...||_{1}$ norm to $Q_{Ai}:L^{2}(w)\rightarrow
L^{2}(w)$. where $Q_{Ai}$ is defined in (\ref{QA}). Using (\ref{as_De}), (\ref%
{QA}), and (\ref{||uv||}) we have
\begin{eqnarray}  \label{TW35}
&&\hspace{-1cm}||\mathcal{K}_{n}-Q_{Ai}||_{1}\leq \\
&&\frac{1}{2}\int_{0}^{\infty }\left( ||\varphi _{n}(\cdot +z)-Ai(\cdot
+z)||_{L^{2}(w)}+||\psi _{n}(\cdot +z)-Ai(\cdot +z)||_{L^{2}(w)}\right)
\notag \\
&& \times \left( ||\varphi _{n}(\cdot +z)||_{L^{2}(w)}+||\psi _{n}(\cdot
+z)||_{L^{2}(w)}+||Ai(\cdot +z)||_{L^{2}(w)}\right) dz  \notag \\
&& \leq C\int_{0}^{\infty }\left( ||\varphi (\cdot +z)-Ai(\cdot
+z)||_{L^{2}(w)}+||\psi (\cdot +z)-Ai(\cdot +z) ||_{L^{2}(w)}\right)
e^{-z}dz  \notag
\end{eqnarray}%
for some $n$-independent $C>0$. Here we have again used (\ref{as_De}), implying
\begin{equation} \label{TW36}
||\varphi _{n}(\cdot +z)||_{L^{2}(w)}\leq Ce^{-z},\quad ||\psi _{n}(\cdot
+z)||_{L^{2}(w)}\leq Ce^{-z}.
\end{equation}%
Now we can use the dominated convergence theorem to make the limit $%
n\rightarrow \infty $ in (\ref{TW35}).

To pass to the limit $n\rightarrow \infty $  in $\psi _{n}(x)\epsilon \varphi
_{n}(y)$, remark that uniformly in $y>s$
\begin{eqnarray}
&&\epsilon \varphi _{n}(y)=c_{\varphi _{n}}-\int_{y}^{n^{2/3}\gamma(L-2) }\varphi
_{n}(z)dz,\notag\\&&c_{\varphi _{n}}=\frac{1}{2}\int_{-n^{2/3}\gamma(L+2) }^{n^{2/3}\gamma(L-2) }\varphi
_{n}(z)dz=\frac{n^{1/2}}{2}\int_{-L}^{L }\psi^{(n)}_n(\lambda)d\lambda,
\label{ephin}\end{eqnarray}%
where $L$ was defined in Remark \ref{r:1}.
But according to the results of \cite{De-Co:99a},
\begin{eqnarray*}
&n^{1/2}&\int_{-2+n^{-1/4}}^{2-n^{-1/4} }\psi^{(n)}_n(\lambda)d\lambda\to 0,\\
&n^{1/2}&\left(\int_{2+n^{-1/4}}^{L }+\int^{-2-n^{-1/4}}_{-L }\right)\psi^{(n)}_n(\lambda)d\lambda\to 0,\quad
n\to\infty.
\end{eqnarray*}
Thus, (\ref{as_De}) and the evenness of $\psi^{(n)}_n$  yield
\[
\lim_{n\rightarrow \infty }c_{\varphi _{n}}=\lim_{n\rightarrow \infty }n^{1/2}
\int_{2-n^{-1/4} }^{2+n^{-1/4} }\psi^{(n)}_n(\lambda)d\lambda=
\int_{-\infty}^\infty Ai(x)dx=1.
\]
Moreover, (\ref{as_De})
allow us to pass to the limit $n\rightarrow \infty $ in the
second term of the representation (\ref{ephin}) of $\epsilon\varphi_n$. Thus we have uniformly in $y\geq
s>-\infty $%
\begin{equation*}
\lim_{n\rightarrow \infty }\epsilon \varphi _{n}(y)=1-\int_{y}^{\infty
}Ai(z)dz.
\end{equation*}
Now (\ref{||uv||}) implies
\begin{equation*}
\lim_{n\rightarrow \infty }\left( \mathcal{K}_{n}(x,y)+\frac{1}{2}\psi
_{n}(x)\epsilon \varphi _{n}(y)\right) =Q_{Ai}(x,y)+\frac{1}{2}Ai(x)\left(
1-\int_{y}^{\infty }Ai(z)dz\right) ,
\end{equation*}%
where the limit is understood in the $||\dots ||_{1}$-norm.

To prove that $-\partial _{y}\mathcal{K}_{n}:L^{2}(w^{-1})\rightarrow
L^{2}(w)$ converges in the $||...||_{1}$ norm to $-\partial
_{y}Q_{Ai}:L^{2}(w^{-1})\rightarrow L^{2}(w)$, we repeat the argument used
in (\ref{TW35}), taking into account (\ref{||uv||}) with $w_{1}=w^{-1}$ and $%
w_{2}=w$. Besides, we have the relations
\begin{equation}\label{TW37}
||\varphi _{n}(\cdot +z)||_{L^{2}(w^{-1})}\leq C^{\prime }e^{-z},\quad
||\psi _{n}(\cdot +z)||_{L^{2}(w^{-1})}\leq C^{\prime }e^{-z},
\end{equation}%
\begin{equation*}
\lim_{n\rightarrow \infty }||\varphi _{n}(\cdot +z)-Ai(\cdot
+z)||_{L^{2}(w^{-1})}=\lim_{n\rightarrow \infty }||\psi _{n}(\cdot
+z)-Ai(\cdot +z)||_{L^{2}(w^{-1})}=0,
\end{equation*}%
and
\begin{equation*}
\lim_{n\rightarrow \infty }||\varphi _{n}^{\prime }(\cdot +z)-Ai^{\prime
}(\cdot +z)||_{L^{2}(w^{-1})}=\lim_{n\rightarrow \infty }||\psi _{n}^{\prime
}(\cdot +z)-Ai^{\prime }(\cdot +z)||_{L^{2}(w^{-1})}=0.
\end{equation*}%
We obtain then
\begin{equation*}
\lim_{n\rightarrow \infty }\left(-\partial _{y}\mathcal{K}_{n}(x,y)-\psi
_{n}(x)\varphi _{n}(y)\right)=-\partial _{y}Q_{Ai}(x,y)-\frac{1}{2}Ai(x)Ai(y),
\end{equation*}%
where the limit is understood in the $||...||_{1}$ norm.

We are left to prove that $\mathcal{I}_{n}:L^{2}(w)\rightarrow L^{2}(w^{-1})$
converges in the $||...||_{1}$ norm to the operator from $L^{2}(w)$ to $%
L^{2}(w^{-1})$, defined by the kernel
\begin{equation*}
\int \epsilon (x-x^{\prime })Q_{Ai}(x^{\prime },y)dx^{\prime }.
\end{equation*}
To this end denote
\begin{equation*}
\Phi _{n}(x)=\int_{x}^{\infty }\varphi _{n}(x^{\prime })dx^{\prime },\quad
\Psi _{n}(x)=\int_{x}^{\infty }\psi _{n}(x^{\prime })dx^{\prime }.
\end{equation*}%
Then
\begin{equation*}
\begin{array}{rcl}
\epsilon \varphi _{n}(x) & = & \dfrac{1}{2}\displaystyle\int_{-\infty
}^{\infty }\varphi _{n}(x^{\prime })dx^{\prime }-\Phi _{n}(x)=c_{\varphi
_{n}}-\Phi _{n}(x), \\
\epsilon \psi _{n}(x) & = & \dfrac{1}{2}\displaystyle\int_{-\infty }^{\infty
}\psi _{n}(x^{\prime })dx^{\prime }-\Psi _{n}(x)=-\Psi _{n}(x) \\
\epsilon \mathcal{K}_{n}(x,y) & = & \dfrac{1}{2}\displaystyle%
\int_{0}^{\infty }dz(\epsilon \psi _{n}(x+z)\varphi _{n}(y+z)+\epsilon
\varphi _{n}(x+z)\psi _{n}(y+z)) \\
& = & -\dfrac{1}{2}\displaystyle\int_{0}^{\infty }dz(\Psi _{n}(x+z)\varphi
_{n}(y+z)+\Phi _{n}(x+z)\psi _{n}(y+z))+\dfrac{c_{\varphi_n}}{2}\Psi _{n}(y).
\end{array}%
\end{equation*}%
Here the second relation follows from the fact that $\psi_{n-1}^{(n)}$ is an odd function,
and the third one follows from the first, and the second, combined with (\ref{ephin}).
Hence, repeating again the argument used in (\ref{TW35}) and taking into
account that (\ref{as_De}) implies
\begin{equation*}
\lim_{n\rightarrow \infty }||\Psi _{n}(x)-\int_{x}^{\infty }Ai(x^{\prime
})dx^{\prime }||_{L^{2}(w^{-1})}=\lim_{n\rightarrow \infty }||\Phi
_{n}(x)-\int_{x}^{\infty }Ai(x^{\prime })dx^{\prime }||_{L^{2}(w^{-1})}=0,
\end{equation*}%
and
\begin{equation*}
||\Psi _{n}(\cdot +z)||_{L^{2}(w^{-1})}\leq Ce^{-z},\quad ||\Phi _{n}(\cdot
+z)||_{L^{2}(w^{-1})}\leq Ce^{-z},
\end{equation*}%
we obtain
\begin{eqnarray*}
\lim_{n\rightarrow \infty }\mathcal{I}_{n}(x,y)&=&-\int_{x}^{\infty
}Q_{Ai}(x^{\prime },y)dx^{\prime } \\
&&+\frac{1}{2}\left( \int_{y}^{x}Ai(x^{\prime })dx^{\prime
}+\int_{x}^{\infty }Ai(x^{\prime })dx^{\prime }\int_{y}^{\infty
}Ai(y^{\prime })dy^{\prime }\right),
\end{eqnarray*}
where the limit is understood in the $||\dots ||_{1}$ norm. Thus we have
proved assertion (ii).

Note that we have also proved that $\lim_{n\rightarrow \infty }\widehat{%
\mathcal{K}}_{n}(x,y)=\widehat{Q}_{Ai}(x,y)$ uniformly in $x,y\in (s,\infty
) $. Hence assertion (i) is also proved. $\square $


\small
\begin{thebibliography}{99}

\bibitem{APS:97} Albeverio, S., Pastur, L., Shcherbina, M.: On Asymptotic
Properties of the Jacobi Matrix Coefficients.  Matem. Fizika,
Analiz, Geometriya {\bf 4}, 263-277 (1997)

\bibitem{APS:01} Albeverio, S., Pastur, L., Shcherbina, M.: On the $1/n$
expansion for some unitary invariant ensembles of random matrices.
 Commun. Math. Phys. {\bf 224},  271-305 (2001)

\bibitem{BPS:95} Boutet de Monvel, A., Pastur L., Shcherbina M.: On the
statistical mechanics approach in the random matrix theory.
Integrated density of states.  J. Stat. Phys. \textbf{79},
585-611 (1995)

\bibitem{C-K:06} Claeys, T., Kuijalaars, A.B.J.: Universality of the double scaling limit
in random matrix models, Comm. Pure Appl. Math. \textbf{59}, 1573-1603 (2006)

\bibitem{De-Co:98} P. Deift, T. Kriecherbauer, K. T.- R. McLaughlin: New
results on the equilibrium measure for logarithmic potentials in
the presence of an external field, J. Approx. Theory \textbf{95},
 388--475 (1998)


\bibitem{De-Co:99} Deift, P., Kriecherbauer, T., McLaughlin, K., Venakides,
S., Zhou, X.: Uniform asymptotics for polynomials orthogonal with
respect to varying exponential weights and applications to
universality questions in random matrix theory. Commun. Pure
Appl. Math. \textbf{52}, 1335-1425 (1999)

\bibitem{De-Co:99a} Deift, P., Kriecherbauer, T., McLaughlin, K., Venakides,
S., Zhou, X.: Strong asymptotics of orthogonal polynomials with
respect to exponential weights.  Commun. Pure Appl. Math.
\textbf{52}, 1491-1552 (1999)

\bibitem{De-G:07} Deift, P., Gioev, D.: Universality in random matrix
theory for orthogonal and symplectic ensembles.
Int. Math. Res. Papers.2007; 004-116

\bibitem{De-G:07a} Deift, P., Gioev, D.: Universality at the edge of the spectrum
for unitary, orthogonal, and symplectic ensembles of random matrices
Comm. Pure Appl. Math. \textbf{60},  867-910 (2007)

\bibitem{DGKV:07} Deift, P., Gioev, D., Kriecherbauer, T., Vanlessen, M.:
Universality for orthogonal and symplectic Laguerre-type ensembles.
J.Stat.Phys \textbf{129}, 949-1053 (2007)

\bibitem{Dy} Dyson, D.J.: A Class of Matrix Ensembles.  J.Math.Phys.,\textbf{13}, 90-107 (1972)

\bibitem{Jo:98} Johansson, K.: On fluctuations of eigenvalues of random
Hermitian matrices. Duke Math. J. \textbf{91}, 151-204 (1998)

\bibitem{Go-Kr:69} I. C. Gohberg and M. G. Krein, {\em Introduction to
the Theory of Linear Nonselfadjoint Operators}, AMS, Providence,
1969.

\bibitem{Me:91} M.L.Mehta, M.L.: \emph{Random Matrices}. New York: Academic
Press, 1991

\bibitem{PS:97} Pastur, L., Shcherbina, M.: Universality of the local
eigenvalue statistics for a class of unitary invariant random
matrix ensembles. \emph{J. Stat. Phys.} \textbf{86}, 109-147
(1997)
\bibitem{PS:03} Pastur, L., Shcherbina, M.: On the edge universality of the local
eigenvalue statistics of matrix models.  Matematicheskaya fizika,
analiz, geometriya \textbf{10}, N3, 335-365 (2003)

\bibitem{PS:07} Pastur, L., Shcherbina, M.:
Bulk universality and related properties of Hermitian matrix models
 \textbf{130},  205-250 (2007)

\bibitem{RS}  Reed,M., Simon,B.:\emph{Methods of Modern Mathematical
Physics, Vol.IV}, Academic Press: New York, 1978

\bibitem{Sa-To:97}  Saff, E.,  Totik, V.: \emph{Logarithmic Potentials with
External Fields}. Springer-Verlag, Berlin, 1997

\bibitem{S:05} Shcherbina, M.: Double scaling limit
for matrix models with non analytic potentials.
J. Math. Phys. \textbf{49},  033501-033535 (2008)

\bibitem{S:08} M.Shcherbina. On  Universality  for Orthogonal Ensembles of Random Matrices
Preprint arXiv:math-ph/0701046


\bibitem{St1} Stojanovic, A.:  Universality in orthogonal and
symplectic invariant matrix models with quatric potentials.
Math.Phys.Anal.Geom. \textbf{3},  339-373 (2002)

\bibitem{St2} Stojanovic, A.:  Universalit\'{e} pour des
mod\'{e}les orthogonale ou symplectiqua et a  potentiel quartic.
Math.Phys.Anal.Geom. Preprint Bibos 02-07-98

\bibitem{Tr-Wi:94a}
C.A. Tracy, H. Widom, {\em Level spacing distributions and the Airy kernel},
Comm. Math. Phys. {\bf 159} (1994) 151-174;

\bibitem{Tr-Wi:98} Tracy,  C.A., Widom, H.: Correlation
functions, cluster functions, and spacing distributions for random
matrices. J.Stat.Phys. \textbf{92}, 809-835 (1998)

\bibitem{Tr-Wi:05}
C.A. Tracy, H. Widom, {\em Matrix Kernels for the Gaussian otrhogonal
and symplectic ensembles}, Ann. Ins. Fourier, Grenoble. {\bf 55} (2005) 2197-2207.


\bibitem{Wi:99} Widom, H.: On the relations between orthogonal, symplectic
and unitary matrix models. J.Stat.Phys. \textbf{94}, 347-363 (1999)

\end{thebibliography}
\end{document}